\documentclass{iau} 
\usepackage{graphicx}

\title[Predicting solar cycle from dynamo model] %% give here short title %%
{Predicting a solar cycle before its onset using a flux transport dynamo model}

\author[A.\ R.\ Choudhuri]   %% give here short author list %%
{Arnab Rai Choudhuri$^1$}

\affiliation{$^1$Department of Physics, Indian Institute of Science, \\ Bangalore -- 560012, India
 \\ email: {\tt arnab@iisc.ac.in} }

\pubyear{2017}
\volume{335}  %% insert here IAU Symposium No.
\begin{document}

\maketitle

\begin{abstract}
We begin with a review of the predictions for cycle~24 before its onset.  After
summarizing the basics of the flux transport dynamo model, we discuss how this
model had been used to make a successful prediction of cycle~24, on the assumption
that the irregularities of the solar cycle arise due to the fluctuations in the
Babcock--Leighton mechanism.  We point out that fluctuations in the meridional
circulation can be another cause of irregularities in the cycle.
\keywords{Sun: activity, Sun: magnetic fields, MHD.}
\end{abstract}

\firstsection % if your document starts with a section,
\section{Introduction}

Let us begin with a disclaimer.  This review will focus on the physics of predicting
solar cycles from dynamo models and will refrain from presenting any detailed prediction
for the upcoming cycle~25, which is nowadays becoming a hot topic of research.

\begin{figure}
% \vspace*{-2.0 cm}
\begin{center}
\includegraphics[width=5.1in]{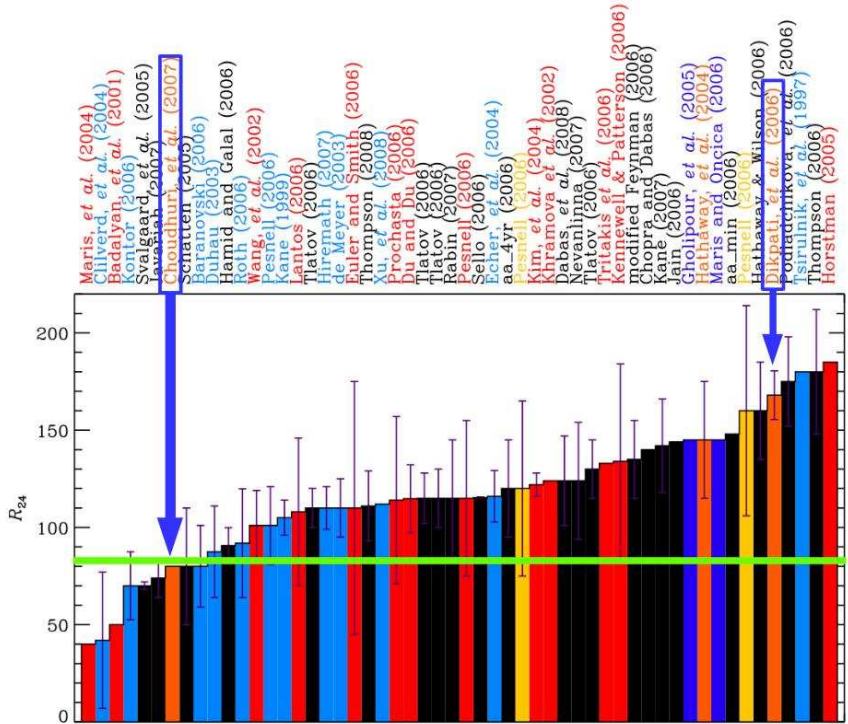} 
%\vspace{5cm}
 \caption{Different predictions of the strength of cycle~24, adopted from Pesnell (2008).
The two predictions based on theoretical dynamo models (Dikpati et al.\ 2006; Choudhuri
et al.\ 2007) are indicated by arrows. The horizontal
line added by us indicates the actual peak strength of cycle~24 reached around April 2014.}
   \label{fig1}
\end{center}
\end{figure}
Now that we know what the present cycle~24 has been like, let us take a look at the
many predictions of cycle~24 before its onset.  Pesnell (2008) produced a plot combining
all the different predictions of the peak sunspot number
of cycle~24.  Figure~1 is adopted from this plot, indicating
the two predictions based on theoretical dynamo models. The first theoretical prediction
by Dikpati et al.\ (2006) was that cycle~24 would be a very strong cycle,
whereas the other prediction by Choudhuri et al.\ (2007) was that it would
be a fairly weak cycle. All the other predictions shown in Figure~1 were
based on various precursors and empirical projections.  We can see that the predictions
covered almost the entire range of possible values of the peak sunspot number from
$\approx$40 to $\approx$190. The horizontal line indicates the actual peak sunspot
number of cycle~24 and was added by us while preparing this presentation. It is clear
that Choudhuri et al.\ (2007) predicted the cycle~24 peak almost correctly.

If several people make several predictions covering the entire possible range, then
somebody's prediction has got to come out right!  Were Choudhuri et al.\
(2007) simply the lucky persons whose prediction accidentally turned out to be correct?
Or did they get it correct because they figured out the correct physics for making such
predictions? We would like to argue that they figured out the correct physics partially,
but not fully. Their success in predicting cycle~24 was due to a combination of
intuition and luck.

In a classic work, Parker (1955) envisaged that the solar cycle is produced by an
oscillation between the toroidal and poloidal magnetic fields of the Sun. Sunspots
form out of the toroidal field due to magnetic buoyancy and provide an indication
of the strength of the toroidal field.  On the other hand, the magnetic fields in 
the polar regions of the Sun are a manifestation of the poloidal field.  We now know
that there is truly an oscillation between these two field components.  The polar
(i.e.\ poloidal) field becomes strongest around the time when the sunspot number (i.e.\
toroidal field) has its lowest value and vice versa. Svalgaard et al.\ (2005)
and Schatten (2005), whose predictions for cycle~24 are included in Figure~1, 
suggested that the polar field at the beginning of a cycle is a good precursor for
the strength of the cycle and used the weak polar field at the beginning of cycle~24
to predict essentially the same low value of the cycle peak that was predicted from the
theoretical dynamo calculations of Choudhuri et al.\ (2007). While 
discussing the physics of cycle prediction, we need to address the question of why the
polar field at the beginning of a cycle acts as such a good precursor of the cycle 
strength. 

\section{Basics of flux transport dynamo model}

The flux transport dynamo model, which started being developed in the 1990s (Wang
et al.\ 1991; Choudhuri et al.\ 1995; Durney 1995; Dikpati \& Charbonneau 1999; 
Nandy \& Choudhuri 2002) and has been recently reviewed by several
authors (Choudhuri 2011, 2014; Charbonneau 2014; Karak et al. 2014a), has emerged
as an attractive theoretical model of the solar cycle.  We describe the basics
of this model in this Section before getting into the question of cycle prediction
in the next Section.

In order to have an
oscillation between the toroidal and poloidal magnetic components, we need to have
mechanisms to generate each component from the other.  The generation of the
toroidal component from the poloidal component due to stretching by differential rotation
is a straightforward process.  To complete the loop, we need a mechanism to generate
the poloidal component back from the toroidal component.  The flux transport dynamo
model assumes that this is achieved by the Babcock--Leighton (BL) mechanism
(Babcock 1961; Leighton 1964), in which
the toroidal field first gives rise to bipolar sunspots which emerge with a tilt
due to the action of the Coriolis force (D'Silva \& Choudhuri 1993) and then the
poloidal field arises from the decay of these tilted bipolar sunspots (Hazra et al.\
2017).

A dynamo model based just on the idea of toroidal field generation by differential
rotation and poloidal field generation by the BL mechanism gives a
poleward propagating dynamo wave---in accordance with what is called the Parker--Yoshimura
sign rule (Parker 1955; Yoshimura 1975).  In order to explain the 
observation that sunspots appear increasingly
at lower latitudes with the progress of the solar cycle, we need an extra mechanism
to reverse the direction of the dynamo wave and make it propagate equatorward.
Choudhuri et al.\ (1995) realized that this extra mechanism is provided by the
meridional circulation (MC) of the Sun, which is observed to be poleward near the surface
of the Sun.  In order to avoid mass piling up in the polar region, there has to
be a subsurface reverse flow towards the equator.  The majority of flux transport
dynamo models assume an one-cell MC with the equatorward flow
at the bottom of the convection zone.  However, Hazra et al.\ (2014) have shown that
the flux transport model can work even with a multi-cell MC structure as long as there is
an equatorward flow at the bottom of the convection zone.  The toroidal field produced
there due to the strong differential rotation discovered by helioseismology is advected
by this flow equatorward, making sure that sunspots appear at lower latitudes with
the progress of the cycle.  The poloidal field generated by the BL
mechanism near the surface is advected poleward by the poleward MC
there---in accordance with observational data.

Perhaps, at this stage, it is a good idea to state clearly what we mean by
flux transport dynamo. We would refer to a solar dynamo model as a flux 
transport dynamo model if it satisfies the following characteristics:
(i) the Babcock--Leighton (BL) mechanism for generating the poloidal field
at the solar surface is included in the model; and (ii) the meridional circulation (MC)
plays an important role by transporting the poloidal field poleward near
the surface and the toroidal field equatorward at the bottom of the 
convection zone.  While different authors in the past might have meant
slightly different things by flux transport dynamo, we believe that the
definition we give here has come to be regarded as the universally accepted
definition of flux transport dynamo at the present time.

Detailed theoretical calculations based on the flux transport dynamo model are broadly in agreement
with observational data pertaining to the solar cycle. A flux transport dynamo code
based on mean field equations tends to give a periodic solution, unless something
special is done to make the solution irregular. After our brief summary
of the flux transport dynamo model, we come to the question as to what causes
irregularities of the cycle and whether an understanding of that will help us
in predicting future cycles.

\section{Fluctuations in the Babcock--Leighton (BL) mechanism}

It has been known for some time that fluctuations in the poloidal field generation
mechanism can cause irregularities in the cycle (Choudhuri 1992).  Choudhuri et al.\
(2007) suggested that the fluctuations in the BL mechanism would be
the main source of irregularities in the dynamo process.  It is not difficult to
understand how the fluctuations in the BL mechanism may arise. This
mechanism depends on the tilts of bipolar sunspot pairs---a pair with a larger tilt
making a more significant contribution to the poloidal field.  The average tilt of
sunspot pairs is given by Joy's law.  However, one finds a distribution of tilt angles
around this average (Stenflo \& Kosovichev 2012)---presumably caused by the effect of
the turbulence in the solar convection zone through which flux tubes rise to form 
bipolar sunspot pairs (Longcope \& Choudhuri 2002).

\begin{figure}
% \vspace*{-2.0 cm}
\begin{center}
\includegraphics[width=2.0in]{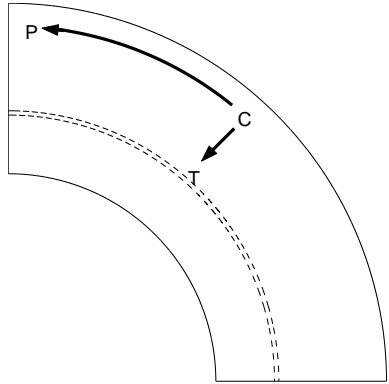} 
% \vspace*{-1.0 cm}
 \caption{A sketch explaining how the correlation between the polar field at the
end of a cycle and the strength of the next cycle arises.  From Jiang et al.\ (2007).}
   \label{fig1}
\end{center}
\end{figure}

Let us first come to the question how the observed correlation between the polar field
at the beginning of a solar cycle and the strength of the cycle arises.  Refer to Fig.~2
taken from Jiang et al.\ (2007), who provided an explanation of this. The BL mechanism
produces poloidal field at C in mid-latitudes near the surface. This poloidal field
will simultaneously be advected poleward by the meridional circulation (MC) to produce
the polar field at P at the beginning of the next cycle and will also diffuse to T at the bottom
of the convection zone, where it will act as the seed of the next cycle. If the fluctuations
in the BL mechanism make the poloidal field produced at C in a cycle stronger than the
usual, then both the polar field at P at the beginning of the next cycle and the seed of the next
cycle at T will be stronger than usual.  On the other hand, if the poloidal field produced
at C is weaker than the usual, then the opposite of this will happen.  We believe that this
is how the correlation between the polar field at the beginning of a cycle and the strength
of the cycle arises.  It may be pointed out that the turbulent diffusion has to be 
sufficiently high to ensure that the poloidal field diffuses from C to T in time of
the order of 5--10 yr, in order to produce the required correlation.  Choudhuri 
et al.\ (2007) used a value of turbulent diffusivity on the basis of mixing length
arguments, which made the correlation come out beautifully.  However, Dikpati et al.\
(2006) used a rather unrealistically low value of turbulent diffusion and the diffusion
time across the convection zone in their model is of the order of 200 yr.  Their model
would not give a correlation between the polar field at the beginning of a cycle and
the strength of the cycle. It may be mentioned that solar dynamo models with high and
low diffusivities have very different characteristics (Jiang et al.\ 2007; Yeates et
al.\ 2008).  There is enough evidence now that models with higher diffusivity are closer
to reality.  Higher diffusivity helps in explaining such observational features as
the dipolar parity (Chatterjee et al.\ 2004; Hotta \& Yokoyama 2010) and the lack of
hemispheric asymmetry (Chatterjee \& Choudhuri 2006; Goel \& Choudhuri 2009).

In order to model actual cycles, one needs to incorporate the actual fluctuations in 
the BL mechanism into the code.  Choudhuri et al.\ (2007) devised a scheme of figuring
out actual fluctuations of the BL mechanism from the observational data of the poloidal fields 
and then incorporating these in the dynamo code.  Since such data are available only from the
1970s, actual cycles could be modelled only from that time. Choudhuri et al.\ (2007)
succeeded in modelling cycles~21--23 reasonably well and cycle~24 was predicted to be
a weak cycle.  Their prediction of cycle~24 was a robust prediction, since they had
incorporated the weakness of the polar field at the beginning of cycle~24 in their
model and the high diffusivity of their model would make this correlated with the
strength of cycle~24.  As we have already pointed out, this prediction has been borne
out triumphantly---making this the first successful prediction of a solar cycle based 
on a theoretical dynamo model in the history of this subject.

\section{Fluctuations in the meridional circulation (MC)}

When Choudhuri et al.\ (2007) made their prediction, it was not realized that the meridional
circulation (MC) probably has occasional large fluctuations and these also may produce
irregularities in the solar cycle. These fluctuations are different from the periodic
variation of MC with the solar cycle, presumably due to the Lorentz
force of the dynamo-generated magnetic field (Hazra \& Choudhuri 2017).
It has been known for some time that the strength
of MC determines the period of the dynamo---faster MC making cycles shorter and vice
versa (Dikpati \& Charbonneau 1999).  
Although we have MC data only for about two decades, we have data pertaining to
durations of cycles for about two centuries and these data indicate that there have
fluctuations in MC with correlation time of the order of a few decades (Karak \&
Choudhuri 2011). What kinds of
irregularities will the fluctuating MC introduce? Suppose MC has slowed down, making cycles
longer. Then diffusion will have more time to act, making cycles weaker.  We may thus
expect longer cycles to be weaker and shorter cycles stronger.  Since a shorter cycle
would imply a faster rise time of the cycle, we may expect a correlation between the
rise time and the strength of a cycle.  Such a correlation is found in the observational
data and is known as the Waldmeier effect.  Karak \& Choudhuri (2011) gave an explanation
of this effect by introducing fluctuations in MC in their model.  

On the basis of such studies, we conclude that there are two important sources of
irregularities in the solar cycle---fluctuations in the BL mechanism and fluctuations
in MC. Choudhuri et al.\ (2007) made their prediction of cycle~24 on the basis of the
assumption that the irregularities are produced by fluctuations in the BL mechanism
alone.  Presumably their prediction was so successful because there had not been large
fluctuations in MC in the last few years. Now that we know the fluctuations of MC to be
another factor introducing irregularities in solar cycles, we need to develop cycle 
prediction methods taking this into account.  We are presently working on this
problem.

It may be mentioned that one big challenge in this field is to develop a theory
of grand minima like the Maunder minimum in the 17th century.  While it has been 
shown that grand minima can be induced by
fluctuations in the BL mechanism alone (Choudhuri \& Karak 2009) or by fluctuations
in MC alone (Karak 2010), we need both of these to develop a comprehensive theory
of grand minima (Choudhuri \& Karak 2012; Karak \& Choudhuri 2013).

\section{Conclusion}

Within the last few years, the flux transport dynamo model has emerged as an attractive
model for explaining the solar cycle and there is increasing evidence that other
solar-like stars also may have similar dynamos working inside them (Karak et al.\ 2014b;
Choudhuri 2017).  It is important that we understand how the irregularities in the
cycle arise, since such an understanding may enable us to predict a future cycle before
its onset. It appears that fluctuations in the BL mechanism and fluctuations in MC are
the two main sources of irregularities in the solar cycle.  Before the beginning of
cycle~24, the role of MC fluctuations was not generally appreciated.  The successful
theoretical prediction of Choudhuri et al.\ (2007) was based on the assumption that
irregularities in the solar cycle are caused only by fluctuations in the BL mechanism.
With the realization that MC fluctuations also can introduce additional irregularities,
it is necessary to develop prediction methods taking this into account.

{\it Acknowledgments.} I thank Gopal Hazra for help in preparing the manuscript. My
research is supported by DST through 
a J.C.\ Bose Fellowship.

\def\apj{{\it ApJ}}
\def\mnras{{\it MNRAS}}
\def\sol{{\it Solar Phys.}}
\def\aa{{\it A\&A}}
\def\gafd{{\it Geophys.\ Astrophys.\ Fluid Dyn.}}


\begin{thebibliography}{}

\bibitem[]{bab61}
  Babcock, H.W. 1961, \apj, 133, 572

\bibitem[]{cha14}
  Charbonneau 2014, {\em ARAA}, 52, 251

\bibitem[]{cha06} 
  Chatterjee, P. \& Choudhuri, A.R. 2006, \sol, 239, 29 

\bibitem[]{cha04}
  Chatterjee, P., Nandy, D., \& Choudhuri, A.R. 2004,  \aa, 427, 1019 

\bibitem[]{cho92}
  Choudhuri, A.R. 1992,  \aa, 253, 277 

\bibitem[]{chou11}
  Choudhuri, A.R. 2011,  {\em Pramana}, 77, 77

\bibitem[]{chou14}
  Choudhuri, A.R. 2014,  {\em Indian J.\ Phys.}, 88, 877

\bibitem[]{chou17}
  Choudhuri, A.R. 2017,  {\em Science China Phys.\ Mech.\ Astron.}, 60, 019601

\bibitem[]{cho07}
  Choudhuri, A.R., Chatterjee, P., \& Jiang, J. 2007, {\em Phys.\ Rev.\ Lett.}, 98, 131103

\bibitem[]{cho09}
  Choudhuri, A.R., \& Karak, B.B. 2009, {\em Res.\ Asron.\ Astrophys.}, 9, 953

\bibitem[]{cho12}
  Choudhuri, A.R., \& Karak, B.B. 2012, {\em Phys.\ Rev.\ Lett.}, 109, 171103

\bibitem[]{cho95}
  Choudhuri, A.R., Sch\"ussler, M., \& Dikpati, M. 1995, \aa, 303, L29

\bibitem[]{sil93}
  D'Silva, S., \& Choudhuri, A.R. 1993,  \aa, 272, 621

\bibitem[]{dik99}
  Dikpati, M., \& Charbonneau, P. 1999, \apj, 518, 508

\bibitem[]{dik06}
  Dikpati, M., de Toma, G., \& Gilman, P.A. 2006, {\em Geophys.\ Res.\ Lett.}, 33, 5102

\bibitem[]{dur95}
  Durney, B.R. 1995, \sol, 160, 213 

\bibitem[40]{goe09} 
  Goel, A., \& Choudhuri, A.R. 2009,  {\em Res.\ Asron.\ Astrophys.}, 9, 115

\bibitem[]{haz17a}
  Hazra, G., \& Choudhuri, A.R. 2017, \mnras, 472, 2728

\bibitem[]{haz17}
  Hazra, G., Choudhuri, A.R., \& Miesch, M.S. 2017, \apj, 835, 39

\bibitem[]{haz14}
  Hazra, G., Karak, B.B. \& Choudhuri, A.R. 2014, \apj, 782, 93

\bibitem[38]{hot10b}
  Hotta, H., \& Yokoyama, T. 2010, \apj, 714, L308

\bibitem[]{jia07}
  Jiang, J., Chatterjee, P., \& Choudhuri, A.R. 2007, \mnras, 381, 1527

\bibitem[]{kar10}
  Karak, B.B. 2010, \apj, 724, 1021 

\bibitem[]{kar11}
  Karak, B.B., \& Choudhuri, A.R. 2011, \mnras, 410, 1503

\bibitem[]{kar13}
  Karak, B.B., \& Choudhuri, A.R. 2013, {\em Res.\ Asron.\ Astrophys.}, 13, 1339 

\bibitem[]{kar14a}
  Karak, B.B., Jiang, J., Miesch, M.S., Charbonneau, P., \& Choudhuri, A.R. 2014a, 
{\em Space Sc.\ Revs}, 186, 561

\bibitem[]{kar14b}
  Karak, B.B., Kitchatinov, L.L., \& Choudhuri, A.R. 2014b, \apj, 791, 59
  
\bibitem[]{lei69}
  Leighton, R.B. 1969,  \apj, 156, 1

\bibitem[]{lon02}
  Longcope, D., \& Choudhuri, A.R. 2002,  \sol, 205, 63

\bibitem[]{nan02}
  Nandy, D., \& Choudhuri, A.R. 2002, {\it Science}, 296, 1671

\bibitem[]{par55}
  Parker, E.N. 1955,  \apj, 122, 293

\bibitem[]{pes08}
  Pesnell, W.D. 2008,  \sol, 252, 209

\bibitem[]{sch05} 
  Schatten, K. 2005, \textit{Geophys. Res. Lett.}, 32, L21106 
 
\bibitem[]{par55}
  Stenflo, J.O., \& Kosovichev, A.G. 2012,  \apj, 745, 129

\bibitem[]{sva05} 
  Svalgaard, L., Cliver, E.W., \& Kamide, Y. 2005,
  \textit{Geophys. Res. Lett.}, 32, L01104

\bibitem[]{wan91}
  Wang, Y.-M., Sheeley, N.R., \& Nash, A.G. 1991,  \apj, 383, 431 

\bibitem[]{yea08}
  Yeates, A.R., Nandy, D., \& Mackay, D.H. 2008, \apj, 673, 544 
  
\bibitem[]{yos75}
  Yoshimura, H. 1975, \apj, 201, 740




\end{thebibliography}
\end{document}